\documentclass[12pt,letterpaper]{article}
\usepackage[a4paper, total={6in, 9in}]{geometry}

\usepackage{graphicx}
\usepackage{helvet}
\usepackage{authblk}
\usepackage{hyperref}
\usepackage[utf8]{inputenc}
\usepackage[T1]{fontenc}
\usepackage{upgreek}
\usepackage{amsmath} 
\usepackage{amssymb} 
\usepackage{orcidlink} 
\usepackage[super,comma,sort&compress]{natbib}

\makeatletter
\renewcommand{\maketitle}{\bgroup\setlength{\parindent}{0pt}
\begin{flushleft}
  \textbf{\@title}
  
  \@author
\end{flushleft}\egroup}
\makeatother

\title{Terahertz Chiral Optics with Ordered Carbon Nanotube Architectures}
\date{}

\author[1,2,\orcidlink{0000-0002-9767-6108}]{Gustavo M.\ Rodriguez-Barrios}
\author[2-4,\orcidlink{0000-0002-2567-3506},**]{Andrey Baydin}
\author[2,3,\orcidlink{0000-0002-2031-4918}]{Jacques Doumani}
\author[1-3,\orcidlink{0000-0002-2331-4526}]{Dasom Kim}
\author[2,3,6,8,\orcidlink{0000-0002-8141-3768}] {Henry O. Everitt}
\author[2-7,\orcidlink{0000-0002-4195-0577},*]{Junichiro Kono}

\affil[1]{Applied Physics Graduate Program, Smalley-Curl Institute, Rice
University, 6100 Main Street, Houston, Texas 77005, USA.}
\affil[2]{Department of Electrical and Computer Engineering, Rice University,
6100 Main Street, Houston, Texas 77005, USA.}
\affil[3]{Smalley-Curl Institute, Rice University, 6100 Main Street, Houston,
Texas 77005, USA.}
\affil[4]{Rice Advanced Materials Institute, Rice University, 6100 Main Street,
Houston, Texas 77005, USA.}
\affil[5]{Department of Materials Science and NanoEngineering, Rice
University, 6100 Main Street, Houston, Texas 77005, USA.}
\affil[6]{Department of Physics and Astronomy, Rice University, 6100 Main
Street, Houston, Texas 77005, USA.}
\affil[7]{Carbon Hub, Rice University, Houston, Texas 77005, USA.}
\affil[8]{U.S. Army DEVCOM Army Research Laboratory-South, Houston, Texas 77005, USA}

\affil[*]{Correspondence: kono@rice.edu}
\affil[**]{Correspondence: baydin@rice.edu}

\begin{document}

\maketitle

\section*{SUMMARY}

Chiral optical terahertz (THz) devices have significant technological implications for telecommunications, spectroscopy, and sensing. Engineering tunable, broadband, and cost-effective THz chiral materials has long been recognized as a challenging endeavor and has been the bottleneck hindering the full exploitation of the THz spectrum. Here, we present an artificial structure based on aligned carbon nanotube films that exhibit a tunable broadband circular dichroism (CD) up to 2.8 degrees. Its behavior is reciprocal. The developed theoretical simulations are in agreement with the experiments and predict a further increase of CD signal up to $\sim$30 degrees when more CNT layers are added.

\section*{KEYWORDS}


Terahertz, Engineered Chirality, Carbon Nanotubes 

\section*{INTRODUCTION}

Chirality, a fundamental property in nature, refers to the geometric characteristic of an object that cannot be superimposed on its mirror image~\cite{Barron2021}. This asymmetry manifests itself in various molecular structures and materials, leading to unique optical properties. One such property is circular dichroism (CD), which describes the difference in absorption strength for left- and right-handed circularly polarized light in chiral substances or structures. While CD spectroscopy has been extensively used in the ultraviolet, visible, and infrared regions, recent advances have extended its application to the terahertz (THz) frequency range, which spans from about 100\,GHz to 10\,THz, corresponding to wavelengths ranging from 3000\,$\upmu$m to 30\,$\upmu$m. Large THz CD signals can open new avenues for biosensing\cite{choi_terahertz_2019,zhang_chiral_2022,wang_automatically_2024, choi_chiral_2022}, material studies \cite{aupiais_chiral_2024, luo_large_2023, zeng_photo-induced_2025,xie_carbon-nanomaterial-enabled_2025}, and communication \cite{xie_phase_2024, rajabalipanah_real-time_2020, zhang_plasmonic_2013, cong_electrically_2019, wang_terahertz_2023, ouchi_terahertz_2025}.

Various platforms have been developed to study and enhance CD in the THz regime, each with its advantages and limitations. The most widely used approach is based on metamaterials~\cite{kan_enantiomeric_2015, wang_terahertz_2022, wang_active_2018,bao_nonvolatile_2022, zhou_terahertz_2012, deglinnocenti_recent_2022, Kim2021}. However, they often rely on complicated and expensive fabrication techniques, and their large CD response is optimized over a narrow spectral range, making broadband applications challenging. Another platform to achieve high CD is gyrotropic materials~\cite{mu_tunable_2019, ju_creating_2023} that absorb one circular polarization and transmit another one, but the operation bandwidth depends on a particular magnetic resonance linewidth. For these devices, light propagation is nonreciprocal due to time-reversal symmetry breaking. 
Lastly, twisted stacks of layered materials have been implemented as a simpler strategy to introduce optical chirality in twisted stacks made of achiral parts~\cite{doumani_engineering_2023,han_recent_2023,wang_automatically_2024,fan_programmable_2025}. This method provides new degrees of freedom, such as the interlayer twist angle, allowing for flexible engineering and tuning of chiroptical responses without changing the material or design~\cite{han_recent_2023}. However, finding a material that exhibits high CD in the THz frequency range for this stacking approach remains a difficult task.

Aligned films of single-wall carbon nanotubes (SWCNTs) films have emerged as a promising and cost‐effective material platform for manipulating THz radiation~\cite{wang_mechanisms_2018}. Several SWCNT-based devices, including linear polarizers~\cite{ren_carbon_2009, ren_broadband_2012,zubair_carbon_2016}, polarization rotators~\cite{baydin_giant_2021, kvitsinskiy_terahertz_2020}, and THz modulators~\cite{burdanova_ultrafast_2021} have been demonstrated. Moreover, aligned films of SWCNTs have been utilized to create chiral devices in the UV-visible wave range~\cite{doumani_engineering_2023,doumani_enabling_2025,xu_giant_2024}. 

There exist different types of SWCNT alignment techniques~\cite{beigmoradi_engineering_2018}, but the controlled vacuum filtration (CVF) method~\cite{he_wafer-scale_2016,doumani_macroscopically_2025,gao_science_2019} is best suited for obtaining films of closely packed SWCNTs. It allows the production of high-density ultrathin films with thicknesses of $\sim$20\,nm with nearly perfect alignment, and the process is scalable and inexpensive, making it suitable for large-scale applications.

Here, we have realized broadband, tunable, cost-effective, and fully reciprocal wafer-scale chiral THz devices fabricated with stacked layers of aligned SWCNT films produced by the CVF method. THz transmission spectroscopy experiments revealed a tunable CD of up to 2.8~degrees for a bilayer stacking case of our proof-of-principle devices. Our electromagnetic simulation results are in agreement with the experiments and predict an improvement of CD signals up to $\sim$30~degrees when more CNT layers are added. These devices tackle a key challenge in advancing THz chiroptical technologies.

\section*{RESULTS}

\begin{figure}
     \centering
     \includegraphics[width=0.9\textwidth]{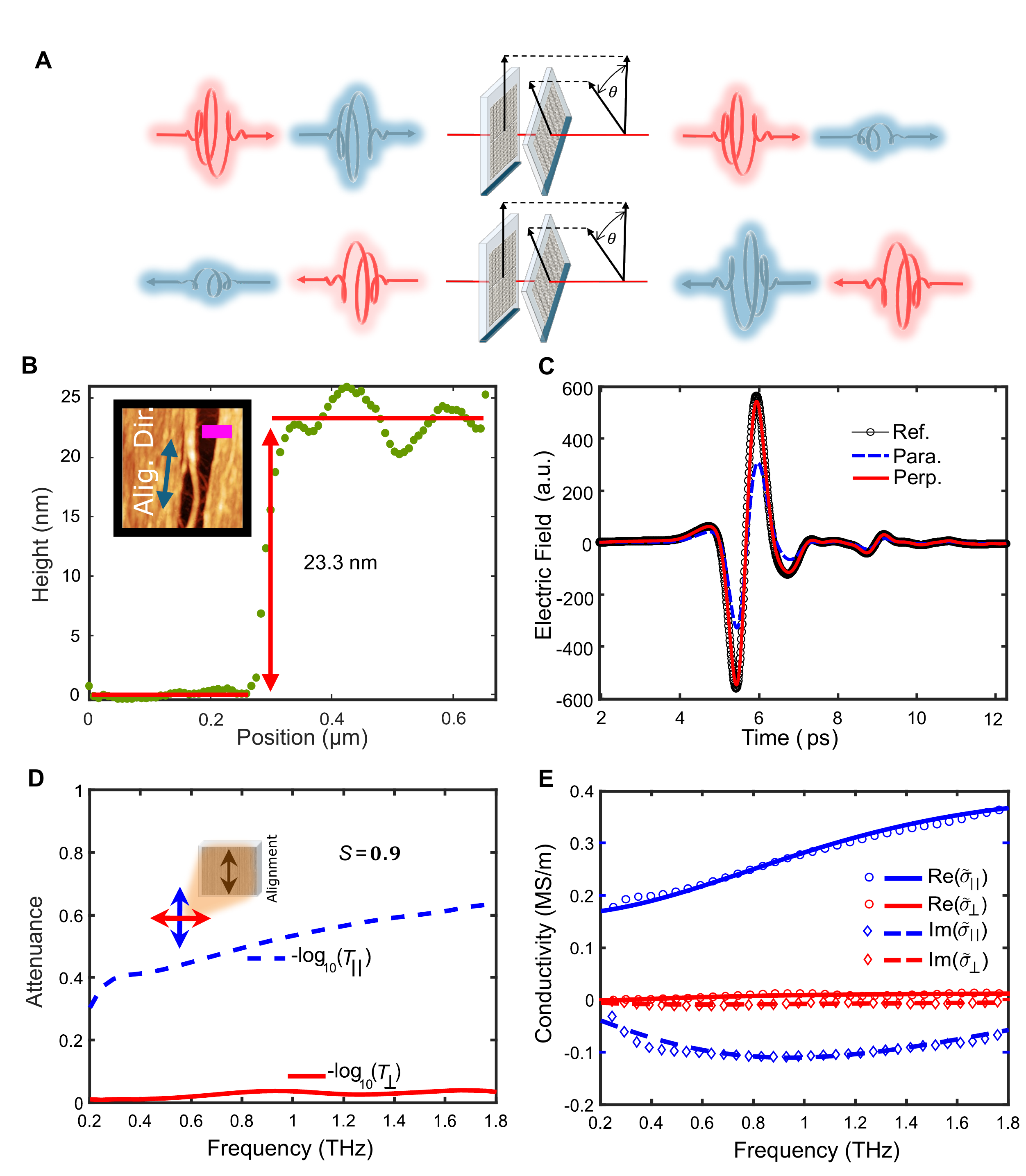}
     \caption{\textbf{Aligned SWCNT films and their THz response.} 
     \textbf{A},~Schematic showing the device structure and the reciprocity of the chiral signal, the chiral response is the same, regardless of whether the light comes from the front or the back. 
     \textbf{B},~Height profile of an aligned SWCNT film, the red line denotes the average thickness of the film; inset: AFM image with the blue arrow indicating the alignment direction, the pink stripe marking the section used for the height profile.
     \textbf{C},~Time-domain THz waveform for the reference (fused silica) and for the SWCNT film for two polarization orientations with respect to the SWCNT alignment direction. 
     \textbf{D},~Polarization-dependent attenuance spectra for aligned SWCNT films. $S$ indicates the nematic order parameter.
     \textbf{E},~Real and imaginary parts of the optical conductivity of the SWCNT film. Markers are the experimental data, and lines are fits to the Drude--Lorentz model. Blue (red) indicates parallel (perpendicular) orientation. Solid and dashed lines are the real and imaginary components, respectively.} \label{Fig:figure 1}
\end{figure}

The idea behind the device is schematically represented in Fig.~\ref{Fig:figure 1}A. Two achiral SWCNT films with strong birefringence and linear dichroism are stacked with twist angle $\theta$. This induces a nonzero CD response, which is also reciprocal to the direction of light propagation. We used the CVF method~\cite{he_wafer-scale_2016} to obtain wafer-scale crystalline SWCNT films in which nanotubes are nearly perfectly aligned (with nematic order parameter $S\sim1$) and maximally packed ($\sim1$ nanotube per cross-sectional area of 1\,nm$^2$). Films were initially fabricated on a porous filter membrane and then transferred to a fused silica substrate; see Methods. Fig.~\ref{Fig:figure 1}B shows an AFM image of a SWCNT film and its height profile. The thickness was determined to be 23\,nm. To access the optical properties of these films, we characterized them using THz time-domain spectroscopy (THz-TDS); see Methods.

Fig.~\ref{Fig:figure 1}C shows time-domain THz waveforms, as measured by THz-TDS, for light polarized parallel ($\parallel$) and perpendicular ($\perp$) to the SWCNT alignment direction and a reference signal for the bare substrate. 
By comparing these traces with the reference signal, it is evident that there is almost no absorption when the electric field is perpendicular to the orientation direction of the SWCNTs, but there is considerable absorption for the parallel case. This is an indication of the high degree of alignment of SWCNTs in these films. The time-domain electric field traces were transformed to the frequency domain by using the standard Fast Fourier Transformation, and the obtained attenuance ($A=-\log_{10}T$) spectra are shown in Fig.~\ref{Fig:figure 1}D. From here, we can estimate the nematic order parameter as $S=(A_\parallel-A_\perp$)/($A_\parallel+A_\perp)=0.9$\cite{imtiaz_facile_2023}, indicating a high degree of alignment. 

For modeling purposes, we extracted the THz optical conductivity of these SWCNT films, which is shown in Fig.~\ref{Fig:figure 1}E. It was obtained by using the thin-film approximation~\cite{tinkham_energy_1956,NussOrenstein}; see Methods. The solid lines are fits to the Drude--Lorentz oscillator model described below~\cite{zhang_plasmonic_2013}. The conductivity is highly anisotropic. 
The model takes into account both the plasmon resonance, which represents the confined collective motion of carriers in each nanotube, and the Drude-like free carrier response describing the intertube transport in the aligned CNT films. The fitting equation for the complex optical conductivity $\tilde{\sigma}$ is~\cite{zhang_plasmonic_2013}
\begin{align}\label{drude-plasmon}
    \tilde{\sigma}(\omega) = \frac{i \sigma_{\text{plasmon}} \omega \gamma_{\text{plasmon}}}{\omega^2 - \omega_0^2 + i \omega \gamma_{\text{plasmon}}} + \frac{i \sigma_{\text{Drude}} \gamma_{\text{Drude}}}{\omega + i \gamma_{\text{Drude}}},
\end{align}
where the first and second terms represent the plasmon and Drude contributions, respectively. From the fits, we obtained the plasmon conductivity at the resonance frequency ($\omega_0/2\pi=3.72\,\text{THz}$) $\sigma_\text{plasmon}=0.3$\,MS/m, the Drude conductivity in the DC limit $\sigma_\text{Drude}\sim 0.2$\,MS/m~\cite{he_wafer-scale_2016}, and the scattering rates $\gamma_\text{plasmon}=74\,\text{THz}$ and $\gamma_\text{Drude}=21\,\text{THz}$.
\subsubsection*{Chirality Engineering}

\begin{figure}
     \centering
     \includegraphics[width=0.95\textwidth]{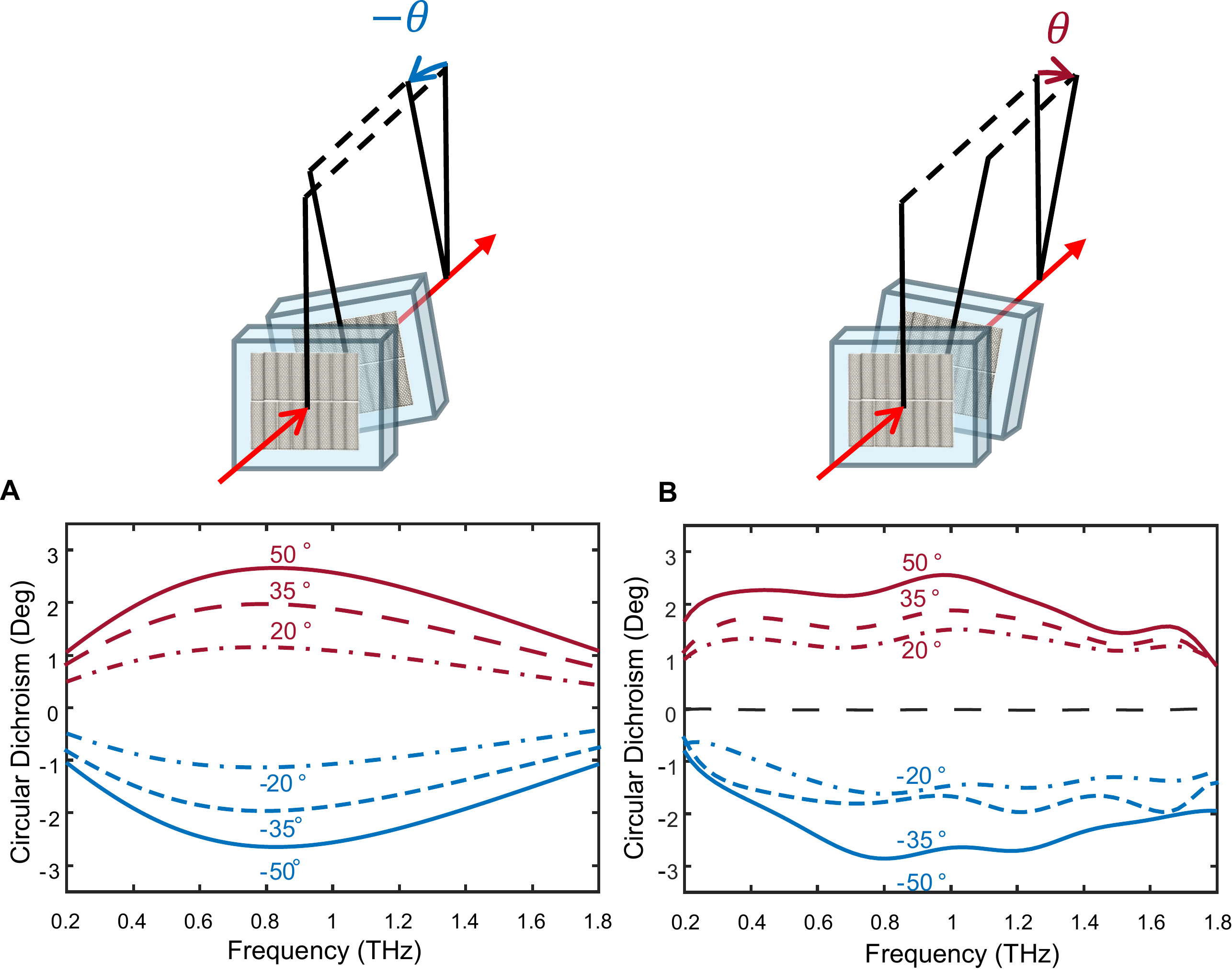}
     \caption{\textbf{Engineered chirality.} 
     \textbf{A},~Simulated CD spectra for the same configuration as \textbf{B}. 
     The schematics show the sign convention used. 
     \textbf{B},~Experimentally measured CD spectra for a twisted bi-layer structure for several twist angles.
     } \label{Fig:figure2}
\end{figure}

By stacking layers of SWCNT films with a separation between each layer, we introduced geometric chirality into the structure. In an $N$ layer configuration, rotating each CNT layer, $i$, by a twist angle $\theta$ with respect to the layer $i-1$ creates an enantiomer whose mirror image is a similar arrangement with $\theta\rightarrow-\theta$. These two structures exhibit distinct chiral properties due to their opposite handedness. To quantify the optical chirality response, we measured the CD, defined as 
\begin{equation}
    \mathrm{CD}=\tan^{-1}\left(\frac{t_\text{r}-t_\text{l}}{t_\text{r}+t_\text{l}}\right),
    \label{eq:CD1}
\end{equation}
 where $t_\text{r}$ and $t_\text{l}$ are the total transmission coefficients for right- and left-handed circularly polarized light~\cite{tranter_circular_2010}. The sketches in Fig.~\ref{Fig:figure2} illustrate the bilayer arrangement and defines the sign of $\theta$. The sign convention used here is that the layer facing the incoming light is set at $\theta=0$, and a clockwise (counterclockwise) rotation of the back layer is considered positive (negative). See Methods for more fabrication details. Fig.~\ref{Fig:figure2}B shows the CD signals for the bilayer configuration with different twist angles. The twist angle controls the signal amplitude, i.e., CD increases as the twist angle goes from $20\deg$ to $50 \deg$. We used the Jones matrix formalism to model the experimental results, and the simulated traces are shown in Fig.~\ref{Fig:figure2}A.  The response is broadband and does not cross zero in the range recorded. The CD spectra for corresponding enantiomers show the same intensity but with opposite signs, and the optical response is the same when the light comes from the front or back sides, showing that the signal is reciprocal; see Fig.~\ref{Fig:figure 1}A. 

Our CNT architectures demonstrate a broadband chiral response. Fig.~\ref{Fig:figure3}A shows the CD intensity as a function of twist angle and frequency. The only regions where the calculated response is zero are the cases when the films are either parallel or perpendicular configuration in terms of nanotube alignment direction. A cut of the color map at 1\,THz is displayed in Fig.~\ref{Fig:figure3}B, indicating that the maximum signal occurs when $\theta\sim60\deg$. The fraction of transmitted electric field for $t_\text{r}$ and $t_\text{l}$ is shown in Fig. S2. Furthermore, the value of CD increases as the number of layers increments. We experimentally fabricated devices for three and four layers and compared their performance with the simulated results, see Fig. S3. Moreover, simulations of multiple aligned SWCNT layers show that the CD signal strengthens with increasing layer count, while the twist angle per layer for maximum CD signal decreases, as illustrated in Fig.\ref{Fig:figure3}D and Fig.\ref{Fig:figure3}C respectively.

While the measurements were done for devices where SWCNT films are separated by the fused silica substrate (1 mm thickness), below we show that the CD intensity is independent of the film separation. Within the Jones matrix formalism, the $k$-th SWCNT layer can be represented as $R^{-1}(\theta_k)tR(\theta_k)$, where
\begin{align}
t=\begin{bmatrix}
t_{xx} & 0 \\
0 & t_{yy}
\end{bmatrix}    
\end{align}
is the complex transmission matrix of the SWCNT film and
\begin{align}
R(\theta)=\begin{bmatrix}
\cos{\theta} & -\sin{\theta} \\
\sin{\theta} & \cos{\theta}
\end{bmatrix}
\end{align}
is the two-dimensional rotation matrix. Since the fused silica substrate has a negligible extinction coefficient in the current case, the contribution of the substrate to each layer is simply a scalar propagation factor, $P_\text{s} = \exp{(j\kappa_0d_\text{s}n_\text{s})}$, with $d_\text{s}$ and $n_\text{s}$ being the thickness and refractive index of the substrate. Then, the transmission matrix $A$ that  represents the total contribution of an array of $N$ SWCNT films and substrates is 
\begin{align}
A = (P_\text{s})^N\displaystyle \prod_{k=1}^{N} R^{-1}(\theta_k)tR(\theta_k),
\end{align}
and each matrix element is given by 
\begin{align}   
A_{n,m}=(P_\text{s})^N\left(\displaystyle \prod_{k=1}^{N} R^{-1}(\theta_k)tR(\theta_k)\right)_{n,m}
\end{align}
with $n,m\in\{x,y\}$. On single transmission measurements, the CD does not depend on the scalar propagation factor, showing that the separation distance of the CNT films does not affect the CD of the entire structure. 
 \begin{figure}
     \centering
     \includegraphics[width=1\textwidth]{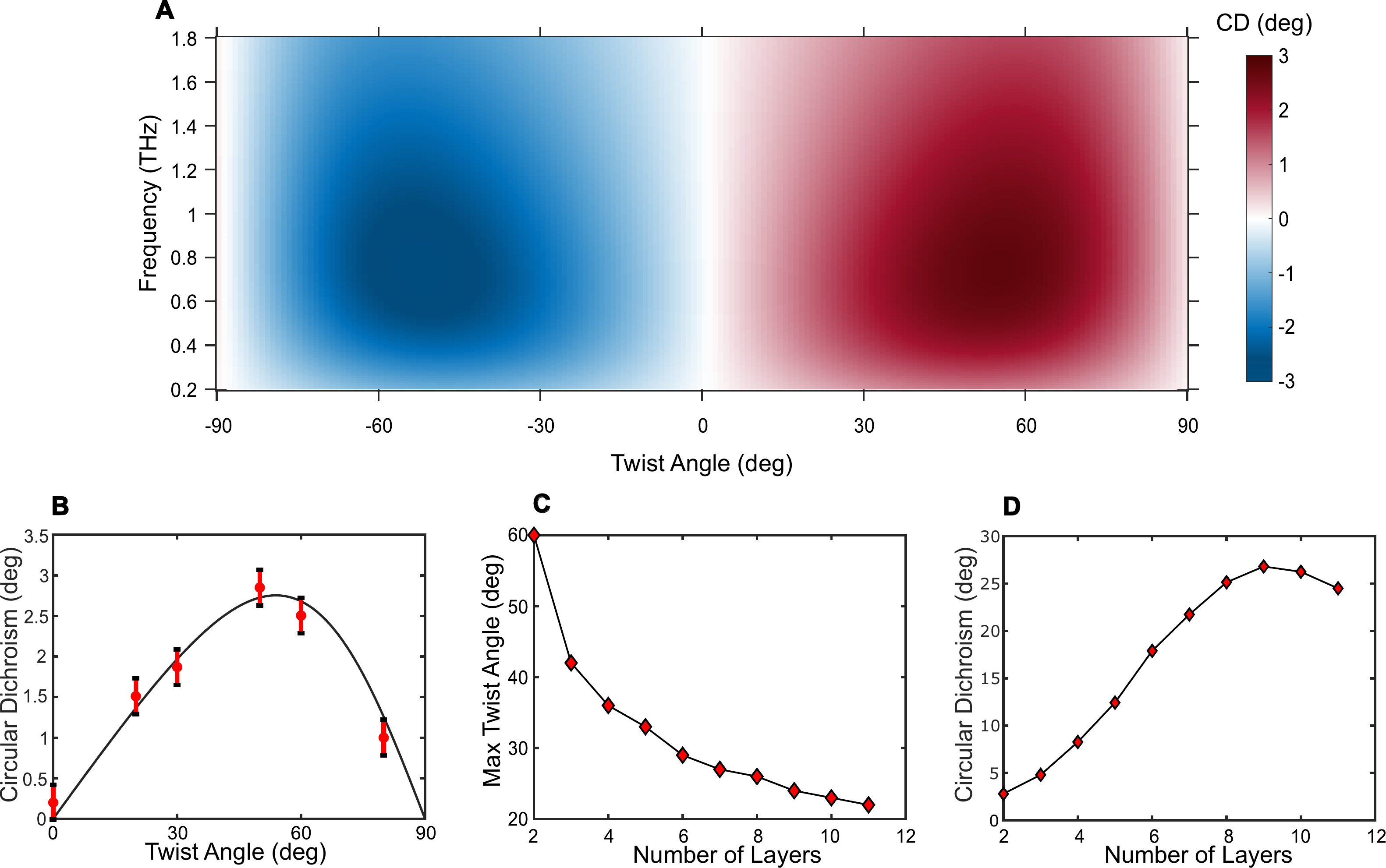}
     \caption{\textbf{CD optimization.}
     \textbf{A},~Color map of calculated CD as a function of frequency and twist angle $\theta$. 
     \textbf{B},~Experimental and simulated CD as a function of the twist angle at 1\,THz. The dots are from experiments. The solid line represents the model, the error bars are the standard deviation for 3 devices. 
     \textbf{C},~Twist angle for maximum CD as a function of number of layers, from simulations. 
     \textbf{D},~Calculated Maximum CD at 1 THz as a function of number of layers. 
     } \label{Fig:figure3}
\end{figure}

\section*{DISCUSSION}

The devices presented here exhibit broadband, reciprocal CD in the THz range, arising from the broadband anisotropy of aligned SWCNT films. Both the real and imaginary parts of the conductivity tensor contribute to the observed response, resulting in finite CD when these films are stacked with a twist angle. This effect would not be expected in analogous assemblies made from wire-grid polarizers, where absorption dominates the response~\cite{yamada_terahertz_2009}.

An important characteristic of this platform is that the observed CD is reciprocal and does not require any external driving mechanism such as magnetic fields, electrical bias, or thermal modulation. This differentiates it from magneto-optical~\cite{mu_tunable_2019, ju_creating_2023} or actively tunable metamaterial approaches~\cite{wang_active_2018}, which often rely on time-reversal symmetry breaking or external fields to produce a chiral response. The ability to generate broadband CD passively simplifies device operation and integration in THz systems.

The CD observed in our devices spans a broad frequency spectrum that includes key THz communication bands, such as 0.3–0.5\,THz (the “G-band”) and 0.85–1.1\,THz. These bands are envisioned as critical for 6G wireless communication and beyond~\cite{akyildiz_terahertz_2022, koenig_wireless_2013, piesiewicz_short-range_2007, sarieddeen_overview_2021}. The ability of our devices to provide polarization control within these frequency ranges directly connects their chiral optical functionality to the requirements of next-generation THz communication technologies, with possible applications as filters or analyzers.

Furthermore, future directions can be directed toward maximizing and tailoring the CD spectral response by employing selective doping or length-controlled enrichment of SWCNT films, which would allow adjustment of the plasmon resonance and the overall anisotropy spectrum.

In summary, stacked SWCNT films prepared by controlled vacuum filtration provide a platform for reciprocal broadband THz CD without external bias. The results presented here establish the basis for tuning and integrating such films into future THz photonic systems.

\section*{METHODS}\label{Methods}
\subsubsection*{Single CNT Film Preparation}
The preparation starts from CNT powder containing metallic and semiconductor SWCNTs, made by the arc-discharge technique, purchased from Carbon Solutions Inc.\ with a product number P2 (purity$>$90wt\%) and an average diameter of 1.4\,nm. Initially, the CNT powder was dispersed in a 1 wt/vol\% sodium deoxycholate (DOC) surfactant solution in purified water at a concentration of 400\,µg/ml. Later, this dispersion was subjected to ultrasonic tip horn sonication for 45 minutes at 21\,W output power, then to remove large bundles, the sonicated dispersion was further purified through ultracentrifugation at 38000 r.p.m. (Sorvall Discovery 100SE Ultracentrifuge using a Beckman SW-41 Ti swing bucket rotor). After the centrifugation, the upper 80\% of the supernatant was collected and further diluted 20-fold in water to be below the critical micelle concentration (CMC $\sim$0.22\,wt/wt\% ). The final weight percent of DOC before film-making was 0.025\,wt/wt\% . In the second step, a vacuum filtration method~\cite{he_wafer-scale_2016} was utilized, where the CNT suspension prepared in the first step is poured into a filtration funnel with a small-pore-size filter membrane, and differential pressure across the filter membrane pushes the suspension slowly through the pores, leaving CNTs on the filter membrane. The pressure was applied in three stages. Initially, the pressure was simply atmospheric, and the flux rate was approximately 22 min/ml (6 drops/minute). After approximately 76 minutes, the rate decreased to 3 drops/minute. A suction pressure (1 inch WC) was applied, setting the rate at 8 drops/minute until all the mix had been filtered. Finally, a pressure of 5 in WC was set to dry the membrane for 7 minutes.  In the end, the film was wet-transferred to a fused silica substrate, and the filter membrane dissolved in chloroform.

\subsubsection*{Terahertz Time-Domain Spectroscopy}
To conduct the transmission terahertz time-domain spectroscopy measurements, we used a commercial TeraFlash Pro system from Toptica Photonics. Using two off-axis parabolic mirrors, the generated THz rays were collimated and further focused into the CNT structures. The diameter of the spot size was $\sim$4\,mm; the transmitted beam was further collimated and focused to the detector by another two off-axis parabolic mirrors.

To calculate the complete complex transmission matrix coefficients $(t_{xx}, t_{xy}, t_{yy},$ $t_{yx})$, we used the standard method~\cite{choi_terahertz_2019}. The emitter and detector had a high degree of linear polarization and were oriented such that the THz beam had the electric field component parallel to the table. Furthermore, three THz wire grid polarizers with extinction ratio of $\sim10^3$ were used in the configuration shown in Fig. S1.  The first and third polarizers (P1 and P3) were oriented to allow the transmission of the component parallel to the table (defined as 0°), and the second polarizer (P2) was mounted in a rotational mount and settled at either +45° or -45°; the sample holder was also in rotational stage allowing us to rotate the sample arbitrarily. 

Since any arbitrary electric field can be expressed as a sum of two orthogonal components, the transmitted(after sample) electric field parallel to the table ($E_x$), can be calculated as $E_x (t)=E_{+45^\circ} (t)+E_{-45} (t)$, where $E_{+45^\circ}$ and $E_{-45}$ are the measured electric fields when P2 is at +45$^\circ$ and -45$^\circ$ respectively. In the same way, the transmitted electric field perpendicular to the table ($E_y$) is $E_y (t)=E_{+45^\circ} (t)-E_{-45^\circ} (t)$. 
The complex frequency-domain electric field spectra were obtained using a fast Fourier transform, $\Tilde{E_j}(\omega)=FFT\{E_j (t)\}$ with $j=x,y$.

The incident (to the sample) beam $\Tilde{E}_\text{in}(\omega)$ is always polarized in the $x$ orientation. It is determined by measuring the bare substrate.\\ When the sample is mounted horizontally, the transmitted light $\Tilde{E}^h(\omega)$ (superscript indicates the orientation of the sample) is
$\Tilde{E}^h(\omega)
=
\begin{bmatrix}
\Tilde{E}^h_x \\
\Tilde{E}^h_y \\
\end{bmatrix}=
\begin{bmatrix}
t_{xx} & t_{yx} \\
t_{xy} & t_{yy} \\
\end{bmatrix}
\begin{bmatrix}
\Tilde{E}_\text{in} \\
0 \\
\end{bmatrix}$. 
Fig. S4 shows the measured raw time domain traces $\Tilde{E}^h_x$, $\Tilde{E}^h_y$, $\Tilde{E}^v_y$, and $\Tilde{E}^v_x$.
Additionally, the transmitted electric field when the sample is oriented vertically is
$\Tilde{E}^v(\omega)
=R(\frac{\pi}{2})
\begin{bmatrix}
t_{xx} & t_{yx} \\
t_{xy} & t_{yy} \\
\end{bmatrix}
R^{-1}(\frac{\pi}{2})
\begin{bmatrix}
\Tilde{E}_\text{in} \\
0 \\
\end{bmatrix}$, 
where $R$ is the rotation matrix. Solving for these equations, the complex transmission coefficients are 
\begin{align}  \label{eq:t_lin}
t_{xx}=\frac{\Tilde{E}^h_x }{\Tilde{E}_\text{in}};\space
t_{xy}=\frac{\Tilde{E}^h_y }{\Tilde{E}_\text{in}};
t_{yx}=\frac{\Tilde{E}^v_y }{\Tilde{E}_\text{in}};
t_{yy}=\frac{\Tilde{E}^v_x }{\Tilde{E}_\text{in}},
\end{align}
The transmission matrix of circularly polarized waves can be calculated as
\begin{align}
T_{\text{cir}} = 
\begin{bmatrix}
t_{rr} & t_{lr} \\
t_{rl} & t_{ll}
\end{bmatrix} = 
\frac{1}{2}
\begin{bmatrix}
t_{xx} + t_{yy} + i(t_{yx} - t_{xy}) & t_{xx} - t_{yy} + i(t_{yx} + t_{xy}) \\
t_{xx} - t_{yy} - i(t_{yx} + t_{xy}) & t_{xx} + t_{yy} + i(t_{xy} - t_{yx}).
\end{bmatrix}
\end{align}
We use the convention that the first subscript denotes the polarization of the incident light and the second denotes that of the transmitted beam.

\subsubsection*{CD measurement}
The total transmission coefficients for any material $t_r$ and $t_l$ are the fraction of light transmitted when the incident light is right and left-handed polarized and are calculated as \cite{choi_terahertz_2019} 
\begin{align} \label{eq:t_cir}
t_r = \sqrt{|t_{rr}|^2 + |t_{rl}|^2} = \frac{1}{\sqrt{2}} \sqrt{|t_{xx} + i t_{yx}|^2 + |t_{xy} + i t_{yy}|^2}, \space \text{and} \\t_l = \sqrt{|t_{ll}|^2 + |t_{lr}|^2} = \frac{1}{\sqrt{2}} \sqrt{|t_{xx} - i t_{yx}|^2 + |t_{xy} - i t_{yy}|^2},
\end{align}
with this, the circular dichroism is calculated as \cite{tranter_circular_2010}
\begin{align}\label{eq:CD}
    \text{CD}=\text{tan}^{-1}\left(\frac{t_r-t_l}{t_r+t_l}\right)
\end{align}

\subsubsection*{Material Parameter Extraction}
Due to the extremely thin nature of the CNT films, the complex conductivity tensor of the films can be calculated using the thin film approximation \cite{tinkham_energy_1956,thoman_nanostructured_2008},
\begin{align} \label{eq:thinfilm}\
    t_{ij}=\frac{1+n_\text{subs}}{1+n_\text{subs}+Z_0\sigma_{ij} d,}
\end{align}
where $t_{ij}$ are the transmission coefficients discussed before, $n_\text{subs}=1.96$ is the refractive index of the substrate, $Z_0\approx377\,\Omega$ is the impedance of free space, and $d$ is the film thickness.
\newpage

\section*{RESOURCE AVAILABILITY}

\subsection*{Data availability}
The data that support the findings of this study are available from the corresponding author upon request.
\section*{ACKNOWLEDGMENTS}

We acknowledge financial support from the Robert A. Welch Foundation (C-1509), the Air Force Office of Scientific Research (FA9550-22-1-0382), and the US National Science Foundation through grant no. PIRE-2230727. 

\section*{AUTHOR CONTRIBUTIONS}

J.K. and A.B. conceived the experiment.  G.R. fabricated the samples, conducted the experiment, analyzed the results, and performed simulations. J.K. and A.B. supervised the project. D.K. and G.R. built the THz-TDS setup, and J.D. prepared the original CNT dispersion and calibrated the CVF setup. G.R. wrote the manuscript. G.R., A.B., H.E. and J.K. edited the manuscript. All authors reviewed the manuscript.

\section*{DECLARATION OF INTERESTS}

The authors declare no competing financial interests.

\section*{DECLARATION OF GENERATIVE AI AND AI-ASSISTED TECHNOLOGIES}

During the preparation of this work, the authors did not use any AI tool and take full responsibility for the content of the publication.

\newpage

\bibliography{references}

\bigskip

\newpage



\renewcommand{\figurename}{\textbf{Figure}}
\renewcommand{\thefigure}{\textbf{S\arabic{figure}}}

\makeatletter


\section*{SUPPLEMENTARY FIGURES}

\begin{figure}[h]
     \includegraphics[width=0.95\textwidth]{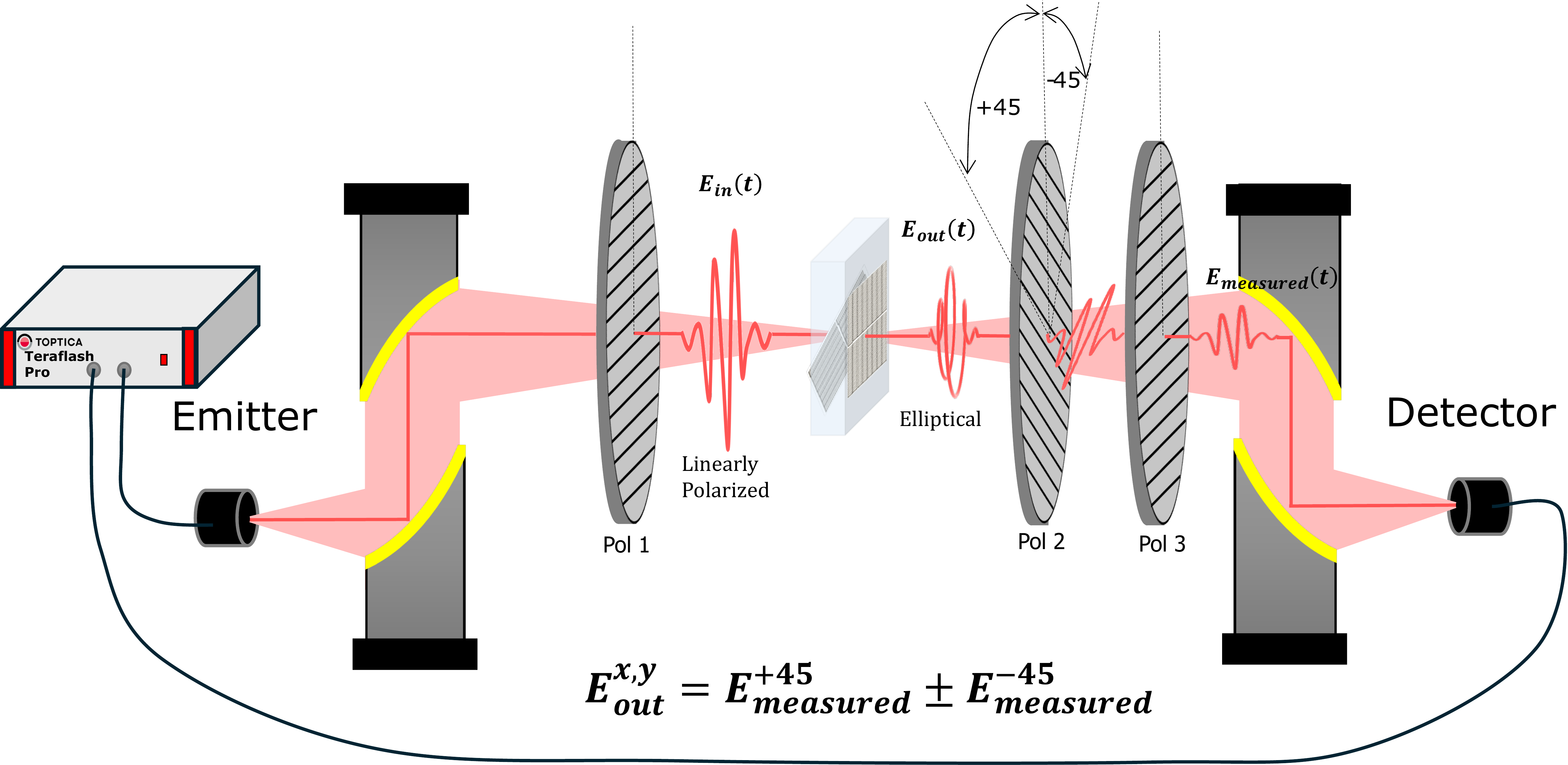}
     \caption{\textbf{THz TDS Experimental Setup}. Two off-axis parabolic mirrors were utilized to collimate and focus the THz beam from the emitter to the sample. The spot size was $\sim$4~mm, and two other off-axis parabolic mirrors were used to collimate and focus the transmitted beam to the detector. To make the measurements of the four transmission coefficients, we utilized three linear polarizers with an extinction ratio of 300:1; the first and third polarizers were aligned to allow the polarization of light parallel to the table, and the second polarizer was the analyzer that was at either plus or minus 45 degrees.  
     } \label{Fig:figure3}
\end{figure}


\begin{figure}[h]
     \includegraphics[width=0.95\textwidth]{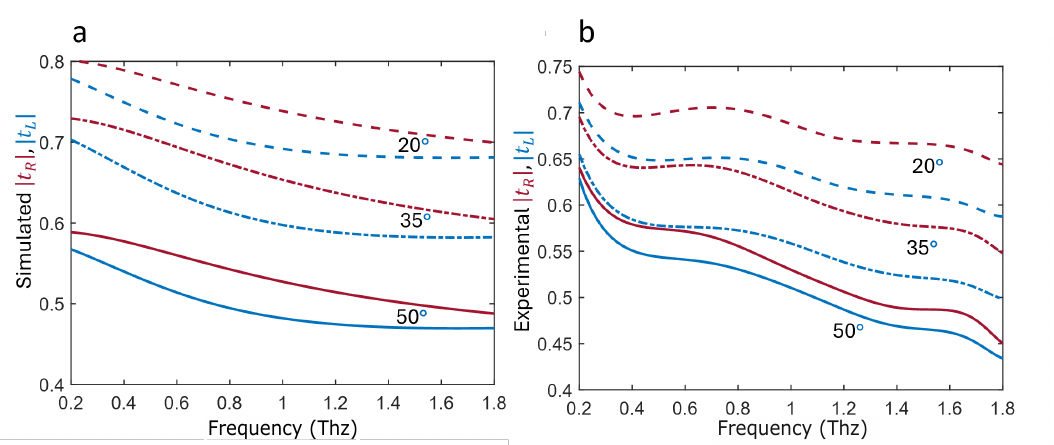}
     \caption{\textbf{Circular Transmission Coefficients}
     Circular transmission coefficients for bilayer chiral structures positive twist angle. The traces correspond to bi-layer devices with twist angles of 20, 35, and 50 degrees. \textbf{a},~Simulated results calculated with the Jones matrix formulation and the optical constants of a single layer. \textbf{b},~ Experimentally measured circular transmission coefficient
     }
\end{figure}


 \begin{figure}[h]
     \centering
     \includegraphics[width=1\textwidth]{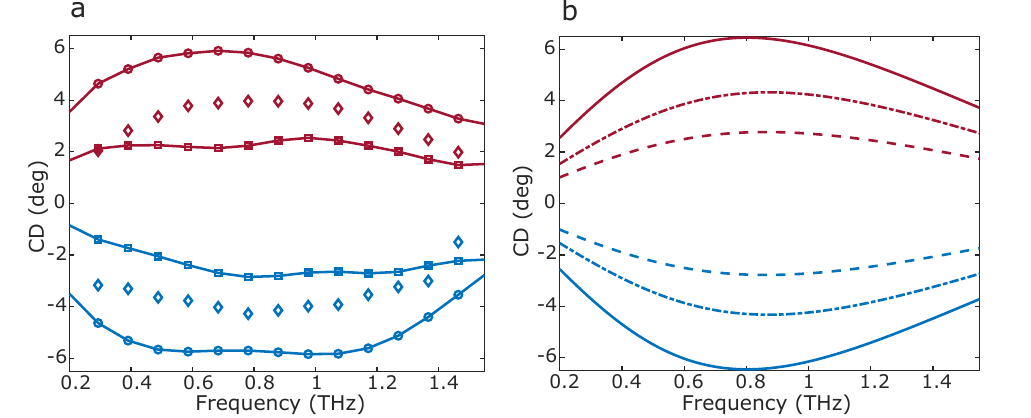}
     \caption{\textbf{2, 3, and 4-layer devices.} 
     \textbf{a},~Experimentally measured CD spectra for devices with 2, 3, and 4 layers, the twist angle was chosen to be 35 degrees for all cases.  
     \textbf{b},~Simulated CD spectra for the same configuration as \textbf{a}. 
     .
     } \label{Fig:CD_multilayers}
\end{figure}


\begin{figure}[h]
     \centering
     \includegraphics[width=1\textwidth]{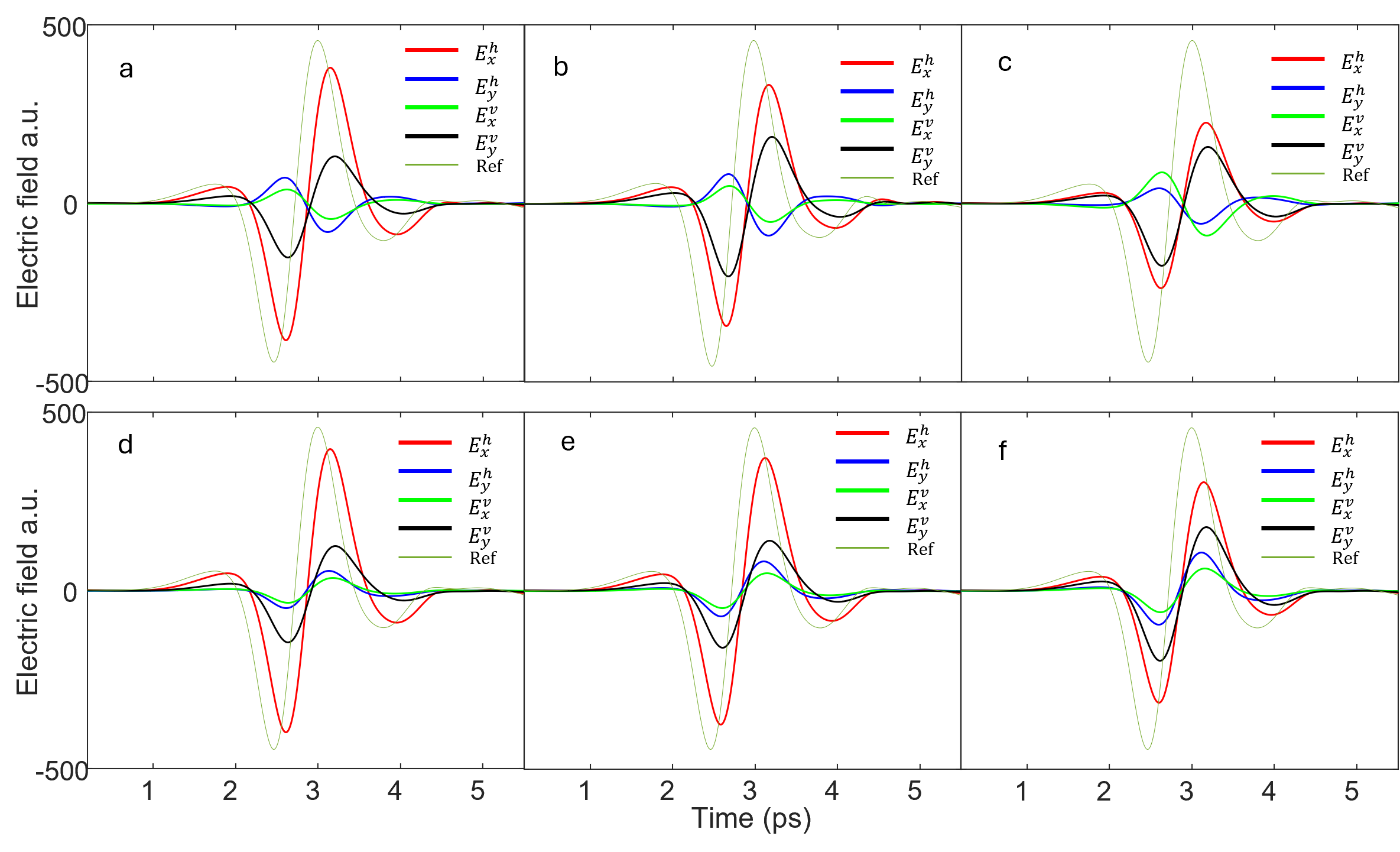}
     \caption{\textbf{Time domain traces}.
     Time domain wave form measured using THz-TDS of engineered chiral samples. Each panel shows all the components $E_x^\text{h},E_y^\text{h},E_x^\text{v}$ and $E_y^\text{v}$ needed to measure the complex transmission coefficients. The traces \textbf{(d,e,f) a, b, c,} correspond to bi-layer devices with twist angle of, for (-)20, (-)35, (-)50 degrees respectively. 
     } \label{Fig:CH3_t_circ}
\end{figure}

\bigskip


\end{document}